\documentclass[prd,aps,preprintnumbers,twocolumn,floatfix,nofootinbib]{revtex4-2}
\usepackage{fix-revtex-4-2-bib}
\usepackage{qcd}
\usepackage{epsfig}
\usepackage{graphicx}
\usepackage{url,hyperref}
\usepackage{bm}
\usepackage{float}

\begin{document}

\title{Avenues for a number density interpretation of dihadron fragmentation functions}

\author{T.~Rogers}
\email{trogers@odu.edu}
\affiliation{Department of Physics, Old Dominion University, Norfolk, VA 23529, USA}
\affiliation{Jefferson Lab, 12000 Jefferson Avenue, Newport News, VA 23606, USA}
\affiliation{\href{https://orcid.org/0000-0002-0762-0275}{ORCID: 0000-0002-0762-0275}}
\author{A.~Courtoy}
\email{aurore@fisica.unam.mx}
\affiliation{Instituto de F\'isica,
  Universidad Nacional Aut\'onoma de M\'exico, Apartado Postal 20-364,
  01000 Ciudad de M\'exico, Mexico}
\affiliation{\href{https://orcid.org/0000-0001-8906-2440}{ORCID: 0000-0001-8906-2440}}
\begin{abstract}
In this comment, we reassess the underlying physics of the number sum rule for dihadron fragmentation functions. We will argue that, currently, there are no settled 
constraints on what constitutes a valid number density interpretation for multihadron fragmentation functions. Imposing overly restrictive criteria might lead to misinterpretating the data. Most importantly, and on the basis of phenomenological analyses, the slightly varying definitions used in previous work are not excluded from possessing legitimate number density interpretations (up to the usual issues with ultraviolet divergences and renormalization), so long as they are paired with appropriate factorization theorems. We advocate for further theoretical analyses to be challenged with experimental data, available at JLab or at the future EIC. 
\end{abstract}
\preprint{JLAB-THY-24-4007}
\date{April 2, 2024}
\maketitle

%%%%%%%%%%%%%%%%%%%%%%%%%%%%%%%%%%%%%%%%%%%%%

Some of the most interesting future uses of both single and multihadron fragmentation functions involve the detailed study of hadron structure. As such, a careful accounting of their interpretations is timely and important.

A recent article~\cite{Pitonyak:2023gjx} criticizes certain types of definitions of dihadron fragmentation functions on the grounds that they do not adequately satisfy a number density interpretation in a field theoretical treatment of QCD. Central to the authors' argument is the \emph{total} multiplicity interpretation of the integral and sum over final states, which appears as  Eq.~(6) and Eq.~(12) of Ref.~\cite{Pitonyak:2023gjx}. Equation~(6) is 
\begin{align}
&\sum_{h_1 \in H} \sum_{h_2 \in H} \int_0^1 \diff{z_2} \int_0^{1-z_2} \diff{z_1} \int \diff[2]{\T{P}{1}} \int \diff[2]{\T{P}{2}} D^{h_1h_2/i} \no
& \qquad = \langle N (N-1) \rangle \, , 
\label{e.eq6}
\end{align}
where $N$ is the total number of hadrons produced from parton $i$ and $D^{h_1h_2/i}$ is a dihadron fragmentation function. 
In Ref.~\cite{Pitonyak:2023gjx}, it is suggested that   formal manipulations of the operators in the multihadron fragmentation function definitions must preserve  Eq.~(\ref{e.eq6}) if a number density interpretation is to remain valid. Furthermore, it is argued that previously used definitions are excluded from having a number density interpretation on this basis. A specific definition is then provided, which is claimed to be capable of proving the validity of this number sum rule. 

As recently shown in Ref.~\cite{Collins:2023cuo}, however, the derivation of this type of sum rule fails due to the orthogonality between the fragmenting quark states and the true asymptotic hadronic final states. The same problem arises already for multiplicity sum rules in %normal 
single hadron fragmentation functions, which appears as Eq.(S23) of Ref.~\cite{Pitonyak:2023gjx}, and is 
%%%%%%%%%%%%%%%%
\begin{equation}
\label{e.hadron_sum_rule}
\sum_{h \in H} \int_0^1 \diff{z} \, d_{h/j}(z) = \langle N \rangle \, ,
\end{equation}
%%%%%%%%%%%%%%%%
with the sum being over all hadron species in a fragmenting quark flavor $j$, $d_{h/j}(z)$ is the ordinary fragmentation function, and $\langle N \rangle$ is the total multiplicity of hadrons in the fragmenting quark.  
Since quark number is strictly conserved in QCD,  the fragmenting quark state $| q \rangle$ has zero overlap with the {\it physical states} that constitute $|X\rangle$.  Whenever those states are chosen to come from the completeness relation involving physical states,  $\sumint |X \rangle \langle X |  = 1$,  
 the assertion that the quantity $\langle N \rangle$  is the expectation value for the total number of hadrons produced by the fragmentation of the quark causes the proof to break down.
For the overlap to satisfy
%%%%%%%%%%%%%%%%
\begin{equation}
\langle q | X' \rangle \neq  0 \, ,
\end{equation}
%%%%%%%%%%%%%%%%
where $|q \rangle$ is the state created by the action of a light-cone quark creation operator on the vacuum,
all final states in $| X' \rangle$ must be included, hence extending the Fock space with a basis for non-hadronic asymptotic states. That is, since the fragmenting quark state $|q \rangle$ has quark number 1, the states in $|X' \rangle$ have quark number 1 or multiples that form hadrons plus 1.
Unlike the physical states involved in Eqs.~(S19) and (S27) of Ref.~\cite{Pitonyak:2023gjx}, that lead to a vanishing overlap with single-particle states, the completeness relation in Reference~\cite{Collins:2023cuo} requires  the inclusion of extra ``orphan'' quarks, to obtain a corrected multiplicity relation.

The result is that a total multiplicity sum rule interpretation is ambiguous by \emph{at least} order 1 hadron (see \cite{Collins:2023cuo} Eq.~(16)) for ordinary fragmentation functions. That ambiguity increases with multihadron fragmentation functions due to the multiple applications of the number operator. The ultraviolet divergences in integrals over transverse momentum exacerbate the ambiguities further. It is important to recognize that the change in the final state Fock space is not a minor modification. It means that the right side of \eref{eq6} cannot be interpreted purely in terms of hadronic state, and the size of the effect cannot be estimated without a detailed treatment of nonperturbative final states.

While the total multiplicity interpretation breaks down, a number \emph{density} interpretation does apply to fragmentation functions, up to the usual issues with renormalization and ultraviolet divergences. This is because constructing a definition for a fragmentation function 
starts with the number operator $\hat{N}_h$ for hadrons of type $h$ with specific momentum, made differential in some choice of variables that are used to characterize the momentum of the final state hadron(s). See, for example, Eq.(12.35) of Ref.~\cite{Collins:2011qcdbook}. Note that changes in variables here do not undermine the number density interpretation. If two different variable choices are related by a simple Jacobian $J$, 
%%%%%%%%%%%%%%%%%
\begin{equation}
\frac{\diff{\hat{N}_h}}{\diff{\Phi}} = J \frac{\diff{\hat{N}_h}}{\diff{\Phi'}}, \label{e.jacobian}
\end{equation}
%%%%%%%%%%%%%%%%%
then $\frac{\diff{\hat{N}_h}}{\diff{\Phi}}$ and $\frac{\diff{\hat{N}_h}}{\diff{\Phi'}}$ both have equally valid number density interpretations in terms of their respective phase spaces, $\diff{\Phi}$ or $\diff{\Phi'}$, as also discussed right after Eq.~(13) of Ref.~\cite{Pitonyak:2023gjx}. At lowest order in perturbation theory, the Jacobian factor can simply be absorbed into the overall hard factor to maintain consistency with a factorization formula. In multihadron fragmentation functions, the setup is the same, but there are multiple applications of the number operator. The multihadron phase space is more complicated, with a wider set of choices for how it may be parametrized and in which variables.  
It is plausible that different choices have practical advantages and disadvantages over one another.
In particular, the factorization theorems are valid only for restricted regions of final state hadron momenta in a complete cross section, and certain choices for parametrizing phase space may be more natural or useful than others when accounting for this. Moreover, the form of the variables will affect the exact form of the evolution kernels when going beyond a pure parton model description. Thus, it is important to maintain some flexibility regarding this choice. However, this is a technical issue rather than one of interpretation. The only concern regarding interpretation is to ensure that the overall factors in the hard contributions of particular factorization formulas are consistent with the Jacobians in transformations like \eref{jacobian}. 
It is also important to recognize that the number density interpretation cannot be taken literally due to the presence of ultraviolet divergences and the need for renormalization. This is especially the case in the connection between transverse momentum dependent and collinear versions of such objections. However, for most ordinary applications they can be treated as if they are number densities, and they retain the properties of a quasiprobability distribution. 

Fortunately, this means that the ability to interpret previously used definitions of dihadron fragmentation functions in terms of number densities is not undermined by the results of Ref.~\cite{Pitonyak:2023gjx}. Such previous definitions include those of the pioneering dihadron work in Refs.~\cite{Bianconi:1999cd,Bacchetta:2002ux} and subsequent phenomenological applications based on it, such as Refs.~\cite{Courtoy:2012ry,Radici:2015mwa}. 

Like most {\it first principle}-based constraints, a misinterpreted sum rule can have practical (numerical) consequences in phenomenological analyses. 
As discussed in 
\cite{Collins:2023cuo,BSA:PAC42}, the semi-inclusive deep inelastic scattering measurements that occur at moderate hard scales, including at facilities like Jefferson Lab, have typical hadron multiplicities around 5, while at a future EIC they are expected to be about 12 to 13~\cite{JLabprivate}. Moreover, only a fraction of the hadrons will lie within the current fragmentation region. Thus, ambiguities of order a few hadrons are significant. 

Therefore, we propose that, rather than appealing to total multiplicity sum rules, the task of refining the precise and optimal definitions for objects like multihadron fragmentation functions, and for determining their evolution, should instead be guided by the requirements for useful factorization theorems. This involves the consideration of the specific regions of final state phase space where the factorization approximations are most accurate, a procedure that becomes increasingly complex with multihadron final states. Likewise, evolution needs to be understood in terms of specific choices for how ultraviolet divergences in transverse momentum integrals are regulated, and which regions of final state phase space they include. 
Already in 2014, a proposal to access multiplicities for dihadron SIDIS at JLab~\cite{BSA:PAC42} complemented the (unpublished) COMPASS measurements~\cite{Makke:2013dya}, which in turn would have been used to test the evolution of dihadron ff to the scales relevant for $e^+e^-$ scattering thanks to Belle data~\cite{Belle:2017rwm}. 
Such proposals could be explored further, both from the theory and phenomology points of view.
Interesting avenues for research lie in this direction.

%%%%%%%%%%%%%%%%%%%%%
\vskip 0.3in
%%%%%%%%%%%%%%%%%%%%%
\acknowledgments
Ted Rogers thanks Elliot Leader for useful discussions. We also thank Elena Boglione, John Collins, Markus Diefenthaler, J.~Osvaldo Gonzalez-Hernandez, Pavel Nadolsky, Jianwei Qiu, Marco Radici and Tommaso Rainaldi for helpful discussions. 
Ted Rogers was supported by the U.S. Department of Energy, Office of Science, Office of Nuclear Physics, under Award Number DE-SC0024715. This work
was also supported by the DOE Contract No. DE- AC05-06OR23177, under which 
Jefferson Science Associates, LLC operates Jefferson Lab. 
Aurore Courtoy was supported  by the UNAM Grant No. DGAPA-PAPIIT IN111222, CONACyT--Ciencia de Frontera 2019 No.~51244 (FORDECYT-PRONACES) and PIIF ``Interconexiones y sinergias entre la física de
altas energías, la física nuclear y la cosmología" (UNAM).
%%%%%%%%%%%%%%%%%%%%%

%%%%%%%%%%%%%%%%%%%%%
\bibliography{bibliography}

%%%%%%%%%%%%%%%%%%%%%
\end{document}